\documentclass[%
 aip,
 jmp,%
 amsmath,amssymb,
 reprint,%
]{revtex4-1}

\usepackage{graphicx}
\usepackage{dcolumn}
\usepackage{bm}
\usepackage{epstopdf}
\usepackage{amssymb}
\usepackage{color}
\usepackage[colorlinks=true, letterpaper=true, pdfstartview=FitV, linkcolor=blue, citecolor=blue, urlcolor=blue]{hyperref}

\begin{document}

\preprint{AIP/123-QED}

\title[]{Promising Ferroelectricity in 2D Group IV Tellurides: a First-Principles Study}

\author{Wenhui Wan}
\thanks{These two authors contributed equally}

\author{Chang Liu}%
\thanks{These two authors contributed equally}

\author{Wende Xiao}%

\author{Yugui Yao}
\email{ygyao@bit.edu.cn}
\affiliation{Beijing Key Laboratory of Nanophotonics and Ultrafine Optoelectronic Systems, School of Physics, Beijing Institute of Technology, Beijing 100081, China}

\date{\today}

\begin{abstract}
Based on the first-principles calculations, we investigated the ferroelectric properties of two-dimensional (2D) Group-IV tellurides XTe (X=Si, Ge and Sn), with a focus on GeTe. 2D Group-IV tellurides energetically prefer an orthorhombic phase with a hinge-like structure and an in-plane spontaneous polarization. The intrinsic Curie temperature $T_{c}$ of monolayer GeTe is as high as 570 K and can be raised quickly by applying a tensile strain. An out-of-plane electric field can effectively decrease the coercive field for the reversal of polarization, extending its potential for regulating the polarization switching kinetics. Moreover, for bilayer GeTe the ferroelectric phase is still the ground state. Combined with these advantages, 2D GeTe is a promising candidate material for practical integrated ferroelectric applications.
\end{abstract}

\maketitle


Nanoscale devices based on ferroelectric thin films and compatible with Si chips have many potential applications, e.g. ultrafast switching, cheap room-temperature magnetic-field detectors, electrocaloric coolers for computers and nonvolatile random access memories.~\cite{Martin2016,Rabe2005,Scott2007} However, it is still a great challenge to keep the ferroelectricity stable in thin films at room temperature to date.~\cite{Kooi2016} For the conventional ferroelectric materials such as BaTiO$_{3}$ or PbTiO$_{3}$, the enhanced depolarization field will destroy the ferroelectricity at critical thicknesses of about 12 \AA\  and 24 \AA,~\cite{Junquera2003,Fong2004} respectively. To address this challenge, new two-dimensional (2D) ferroelectric phases with ferroelectricity sustained against such a depolarization field are desirable to pave the way for the application of "integrated ferroelectrics".

Compared with their bulk counterparts, 2D materials often lose some symmetry elements (e.g. centrosymmetry) as the result of dimensionality reduction,~\cite{Xu2014,Blonsky2015} which favors the appearance of ferroelectricity.
A number of 2D ferroelectric phases have been theoretically proposed, e.g. the distorted $1T$ MoS$_{2}$ monolayer,~\cite{Shirodkar2014} low-buckled hexagonal IV-III binary monolayers including InP and AsP,~\cite{DiSante2015} unzipped graphene oxide monolayer~\cite{Noor-A-Alam2016} and monolayer Group-IV monochalcogenides.~\cite{Wu2016,Fei2016,Wang2016} The Group-IV monochalcogenides including GeS, GeSe, SnS and SnSe have attracted much attention due to their large in-plane spontaneous polarizations $\bm{P_{s}}$ in theory~\cite{Wang2016} and experimental accessibility. Monolayers of SnSe and GeSe have been successfully synthesized.~\cite{Li2013,Sun20141}
However, the ground-state SnSe and GeSe multilayers adopt a stacking order that the directions of $\bm{P_{s}}$ in two neighboring layers are opposite.~\cite{Fei2016} Thus, the non-zero polarization only exists in odd-numbered layers, hindering their ferroelectric applications. Excitingly, a robust ferroelectricity has been experimentally observed in SnTe(001)
few-layers.~\cite{chang2016} Compared with the low Curie temperature $T_{c}=98$ K in bulk SnTe,~\cite{Iizumi1975} the $T_{c}$ in monolayer SnTe was greatly enhanced to 270 K, due to the suppression of the Sn-vacancy and the in-plane expansion of the lattice.~\cite{chang2016, Kooi2016} Meanwhile, the $\bm{P_{s}}$ in 2D SnTe are aligned along the in-plane $<110>$ direction, in contrast to the $<111>$ direction in bulk.~\cite{Plekhanov2014} This behavior again indicates that the dimensionality reduction favors the formation of new ferroelectric phases. However, the ferroelectric properties of 2D SnTe predicted in theory are inconsistent with the experimental measurements.~\cite{chang2016} This calls for a microscopic understanding of the relevant physics.
Additionally, the discovery of the ferroelectricity in 2D SnTe shades light on the possible ferroelectric phase in other Group-IV tellurides. Bulk GeTe is ferroelectric and exhibits a rhombohedral crystal structure.~\cite{DiSante2013} No crystal phase has been identified in bulk SiTe, but several thermodynamically stable phases of 2D SiTe have been theoretically proposed.~\cite{Chen2016} To date, the ferroelectric properties of both 2D GeTe and 2D SiTe remain unexplored.

In this work, we investigated the structural, electronic and ferroelectric properties of 2D Group-IV tellurides XTe (X=Si, Ge and Sn). All computational
details are in the supplementary material ($SM$).~\cite{support} We found that 2D Group-IV tellurides prefer a hinge-like structure with an in-plane spontaneous polarization. During the reversal of polarization, monolayer SiTe undergoes a semiconductor-to-metal transition, while monolayer GeTe and SnTe keep semiconducting and ferroelectric.
Monolayer GeTe has a high ferroelectricity transition $T_{c}$ of 570 K. However, for 2D SnTe, the achievement of the room-temperature ferroelectricity requires external strain. The ferroelectricity of 2D GeTe can be effectively controlled through the application of strain engineering and vertical electric fields. Moreover, bilayer GeTe exhibits a ferroelectric ground state that the polarization of each layer is aligned parallel. These novel properties provide 2D GeTe a promising material for future nanoscale ferroelectric applications.

\begin{figure}[tbp!]
\centerline{\includegraphics[width=0.45\textwidth]{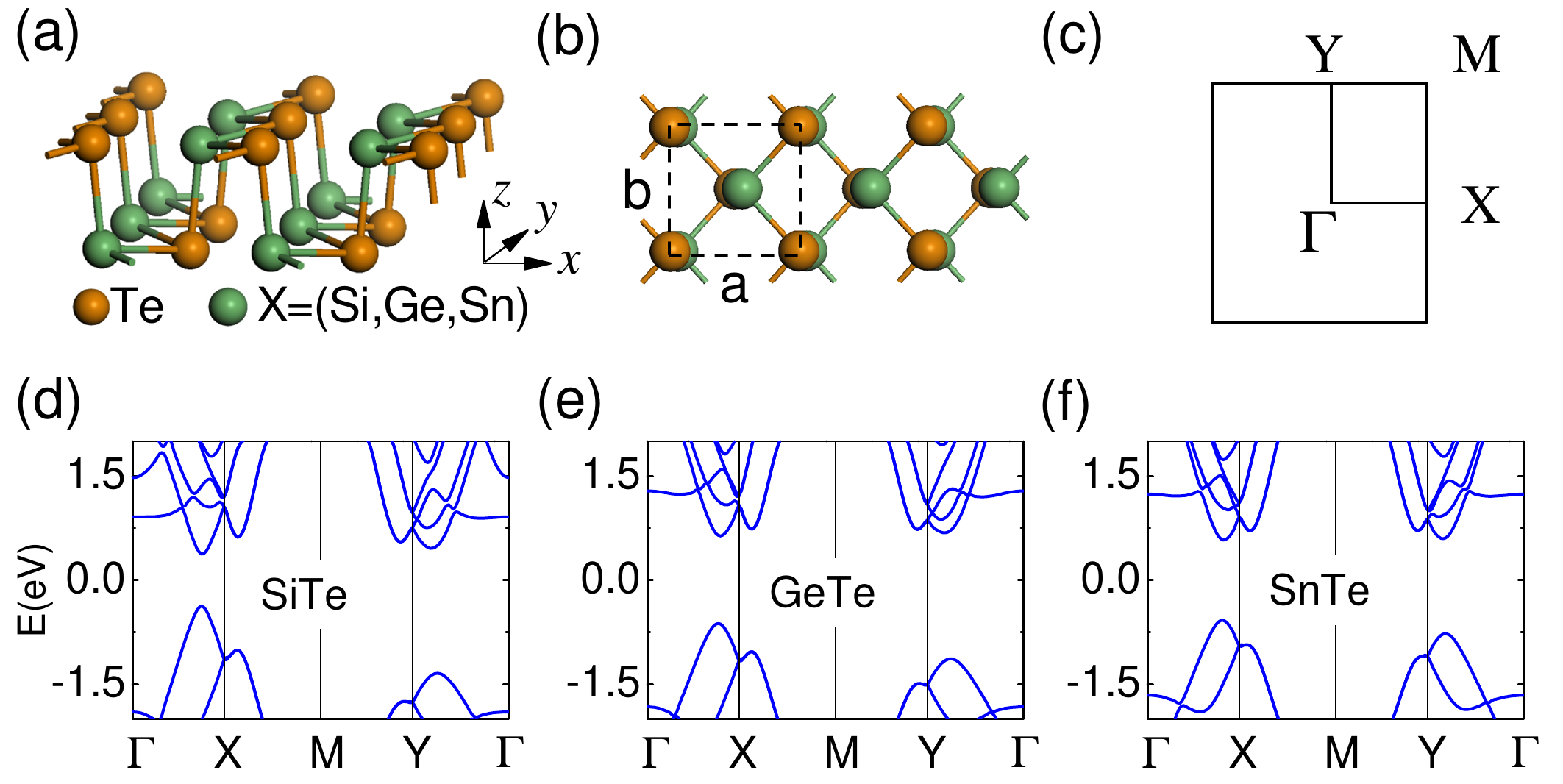}}
\caption{(a, b) Side and top views of the crystal structure of monolayer XTe (X=Si, Ge and Sn), respectively. (c) 2D first Brillouin zone. (d-f) Band structures of monolayer SiTe, GeTe and SnTe, respectively.}
\label{wh1}
\end{figure}

Though bulk GeTe and SnTe adopt a rhombohedral structure below the Curie temperature,~\cite{Iizumi1975,DiSante2013} the synthesized 2D SnTe exhibits a layered orthorhombic phase with a hinge-like structure.~\cite{chang2016} The corresponding monolayer is displayed in Fig.~\ref{wh1}(a) and ~\ref{wh1}(b). Each of the four atoms in a unit cell is three-fold coordinated with the atoms of the other species. $\bm{a}$ and $\bm{b}$ are the lattice vectors along the $x$ (puckered) and $y$ (zigzag) directions, respectively.
We first calculated the crystal structure of the monolayer SnTe with different methods to ensure the reliability of our simulations.
The theoretical results as well as the experimental data~\cite{chang2016} are listed in the Table S1 of $SM$.~\cite{support} In the special hinge-like structure of monolayer SnTe, some atoms are close to each other without covalent bonding (see Fig.~\ref{wh1}(a)). The distance between them is larger than 3.2 \AA.
Description of this kind of interaction due to the weak wave function overlap should include the vdW interactions, which have been reported in the phosphorene.~\cite{qiao2014high} Considering the lattice anisotropy which is critical in the determining the ferroelectricity, it was found that the optPBE-vdW method~\cite{vdw2} produces the best results (see Table S1~\cite{support}) compared with the experiment.

Based on this method, the optimized lattice constants of monolayer Group-IV tellurides are given in Table~\ref{Table1}. No imaginary frequency is observed in the phonon dispersions (see Fig. S1~\cite{support}), confirming their structural stability. Additionally, it is found the orthorhombic phase of monolayer Group-IV tellurides is more stable than the hexagonal phase extracted from the bulk (see the Fig. S2 and Table S2 of $SM$~\cite{support}), in line with the previous theoretical work~\cite{Singh2014} and the experiment.~\cite{chang2016}

\begin{table}[b!]
\caption{\label{Table1}
The lattice constants ($a$, $b$), lattice anisotropy $\delta=(\frac{a}{b}-1)\times100\%$, intrinsic polarization $P_{s}$ and ferroelectric transition barrier $E_{b}$ of monolayer SiTe, GeTe and SnTe.}
\begin{ruledtabular}
\begin{tabular}{lcccccc}
		&$a$ (\AA)& $b$ (\AA) & $\delta$(\%)& $P_{s}({\mu}$C/cm$^{2})$ & $E_{b}$(meV/f.u.)\\
\hline
   SiTe &  4.452  &	4.127     &	7.88  &42.0 &88.5  \\
   GeTe	&  4.472  &	4.273     &	4.66  &32.8 &37.4  \\
   SnTe	&  4.666  & 4.577     & 1.95  &19.4 &4.48  \\
\end{tabular}
\end{ruledtabular}
\end{table}

The calculated band structures of monolayer Group-IV tellurides are shown in Figs.~\ref{wh1}(d-f). The trend of the band gaps is $E_{g}$(GeTe)$>E_{g}$(SnTe)$>E_{g}$(SiTe). The anomalous order of $E_{g}$ might be the result of the fine balance between the relative atomic energy levels and the repulsion between the levels~\cite{Wei1997} The anisotropy of the band structure and phonon dispersion decreases from monolayer SiTe to monolayer SnTe, consistent with the decrease of the anisotropy of their lattice constants (see Table~\ref{Table1}).

\begin{figure}[tbp!]
  \centerline{\includegraphics[width=0.45\textwidth]{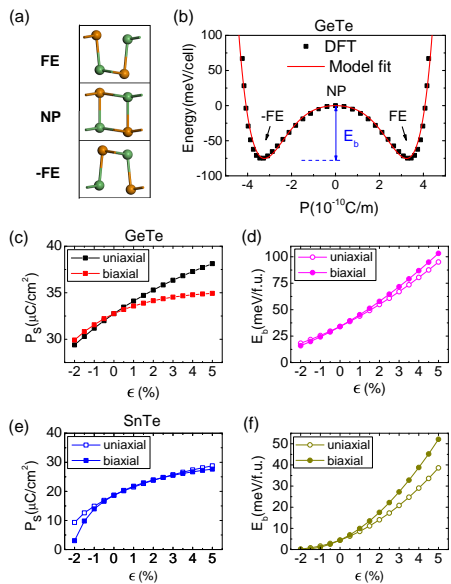}}
  \caption{(a) Two symmetry-equivalent ferroelectric states with opposite in-plane polarizations as well as a high-symmetry, non-polar transition state. (b) Double-well potential of monolayer GeTe vs polarization. $E_{b}$ is the ferroelectric transition barrier. The red line represents the fitting curve of the Landau-Ginzburg model. The uniaxial and biaxial strains dependence of $P_{s}$ and $E_{b}$ for (c, d) monolayer GeTe and (e, f) monolayer SnTe, respectively.}
  \label{wh2}
\end{figure}
In such an orthorhombic structure, the Group-IV atoms displace along the $x$-direction with respect to the Te atoms, leading to the break of the centrosymmetry but the perseveration of the $yz$-mirror symmetry.
Spontaneous polarization is aligned along the $x$-direction and can be labeled as a scalar $P_s$.
Therefore, the thickness of 2D ferroelectric Group-IV tellurides will not be limited by the aforementioned depolarization field vertical to the slab,~\cite{Junquera2003,Fong2004} but another lateral critical size still exists due to the in-plane depolarization field.~\cite{Wang2016} The polarization $P_s$ was calculated by the Berry phase approach.~\cite{King-Smith1993} The reversal of polarization is realized through a phase transition between two symmetry-equivalent ferroelectric states with opposite $P_{s}$ (labeled as the FE state and -FE state in Fig.~\ref{wh2}(a)). By calculating the transition barrier $E_{b}$ of several pathways using the nudged-elastic-band (NEB) methods,~\cite{Henkelman2000} we found that a transition path through a centrosymmetric non-polar (NP) state (see Fig.~\ref{wh2}(a)) has the lowest $E_{b}$. With this transition path, we calculated the polarization dependence of free energy $F(P)$ and show the result specific to monolayer GeTe in Fig.~\ref{wh2}(b). The $P_s$ in the FE state is $3.28\times10^{-10}$ C/m, equivalent to a bulk polarization of $32.8$ ${\mu}$C/cm$^{2}$ if an effective thickness of 1 nm for monolayer GeTe is used. The transition barrier $E_{b}$ is estimated by the energy difference between the FE and NP states (see Fig.~\ref{wh2}(b)). For monolayer GeTe, $E_{b}$ is 74.8 meV, equivalent to 37.4 meV per formula unit (f.u.) and much smaller than $E_{b}\approx 200$ meV/f.u. in conventional ferroelectric PbTiO$_{3}$.~\cite{Cohen1992} This small $E_{b}$ indicates that the required electric field for the reversal of polarization in monolayer GeTe would be much lower than that in PbTiO$_{3}$.

The ferroelectricity can be substantially affected by the external strain.~\cite{Haeni2004} Here the strain is defined as $\epsilon=(\frac{a}{a_{0}}-1)\times100\%$ where $a$ and $a_{0}$ are the lattice constants along the $x-$ or $y-$direction for the strained and unstrained structures, respectively.
It is found that both uniaxial ($\epsilon_{x}$) and biaxial ($\epsilon_{x}=\epsilon_{y}$) tensile strains can enlarge the displacement of the Ge atoms with respect to the Te atoms and therefore effectively enhance the $P_{s}$ and $E_{b}$ of monolayer GeTe, as shown in Figs.~\ref{wh2}(c, d).
In contrast, the compressive strain suppresses the $P_{s}$ and $E_{b}$.

Monolayer SnTe has a larger $P_{s}$ and higher $E_{b}$ than that of GeTe (see Table~\ref{Table1}). However, its band gap calculated using HSE06 approach~\cite{Paier2006} will be closed during the reversal of $P_s$ (see Fig. S3~\cite{support}), leading to a drop of $P_s$ to zero. This semiconductor-metal transition hinders the ferroelectric application of 2D SiTe, but make 2D SiTe suited for the field effect switching devices.~\cite{Metal-Insulator}

Monolayer SnTe remains semiconducting during the reversal of $P_s$. The effective $P_{s}$ and the $E_{b}$ of monolayer SnTe is small (see Table~\ref{Table1}). After applying a tensile strain, both $P_{s}$ and $E_{b}$ can be effectively increased (Figs.~\ref{wh2}(e, f)), showing that the ferroelectricity of 2D SnTe can be effectively tuned by strain.

\begin{figure}[tbp!]
\centerline{\includegraphics[width=0.5\textwidth]{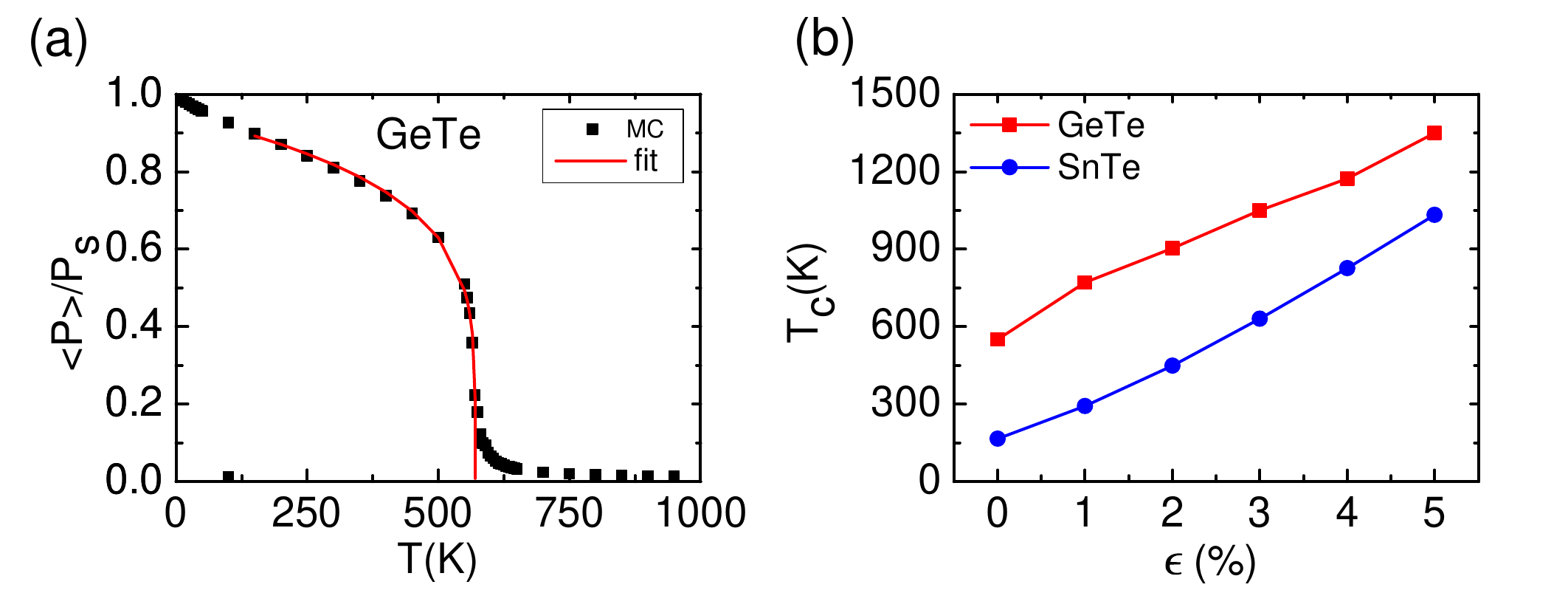}}
\caption{(a) Temperature dependence of averaged polarization $\langle P_{i} \rangle$ of monolayer GeTe obtained by the MC simulations. Here the $\langle P_{i} \rangle$ at different temperature has been normalized with respect to the $\langle P_{i} \rangle_{T=0 K}=P_{s}$. The red line represents a fitting curve of $P_{s}$ with an asymptotic form in the vicinity of the Curie temperatures $T_{c}$. (b) Biaxial strain dependence of $T_{c}$ for monolayer GeTe (red line) and SnTe (blue line).}
\label{wh3}
\end{figure}

The stability of ferroelectricity is represented by the Curie temperatures $T_{c}$ at which the macroscopic spontaneous polarization vanishes. Based on the Landau-Ginzburg phase transition theory,~\cite{Cowley1980,Fei2016} the free energy of GeTe supercell is written as a Taylor expansion in terms of the polarization:

\begin{equation}\label{eq1}
F=\!\sum_{ i } \left( \frac{A}{2}P_{i}^{2} \!+\! \frac{B}{4}P_{i}^{4}\!+\!\frac{C}{6}P_{i}^{6} \right) \!+\!\frac{D}{2}\!\sum_{ < i,j > } \!({P_i} - {P_j})^{2},
\end{equation}

\noindent where $P_{i}$ is the polarization of each unit cell. The first three terms describe the anharmonic double-well potential in a unit cell (see Fig.~\ref{wh2}(b)). The last term represents the dipole-dipole interaction between the nearest neighboring unit cells. The parameter $D$ can be estimated by a fitting process in the mean-field approximation.~\cite{Fei2016} All the fitted parameters are given in the Table S3 of $SM$.~\cite{support} The Monte Carlo simulations were performed with the effective Hamiltonian of Eq.~\ref{eq1} to investigate the ferroelectric phase transition. As shown in Fig.~\ref{wh3}(a), the $T_{c}$ of unstrained monolayer GeTe is 570 K. By applying a biaxial strain of 2\%, the $T_{c}$ can be easily enhanced to 903 K, as illustrated in Fig.~\ref{wh3}(b), consistent with the increase of the transition barrier $E_{b}$ (see Fig.~\ref{wh2}(d)).

In contrast, the $T_{c}$ of unstrained monolayer SnTe is only 166 K (see Fig.~\ref{wh3}(b)), much smaller than the experimental value of $T_{c} = 270$ K.~\cite{chang2016} However, the $T_{c}$ can be increased quickly by applying a biaxial tensile strain (see Fig.~\ref{wh3}(b)). The sensitive response of the ferroelectricity to external strain in 2D SnTe offers a possible reason to explain the difference between the predicted $T_{c}$ and experimental one.
The theoretical lattice anisotropy $\delta$ of monolayer SnTe is $1.95\%$ (see Table.~\ref{Table1}) which is smaller than the experimental $\delta=3.15\%$.~\cite{chang2016} If an uniaxial tensile strain of $\epsilon_{x}=1.0\%$ is applied to monolayer SnTe, the $T_{c}$ can be enhanced from 166 K to 265 K, close to the experimental $T_{c}$ of 270 K.~\cite{chang2016} The external strain has also
been observed in other materials in the vdW epitaxy,~\cite{cai2016band,wang2016band,zhang2015self} such as GaSe flakes via vdW epitaxy on the Si (111) surface.~\cite{cai2016band} The origin of strain in 2D SnTe calls for a future study.

The averaged polarization $\langle P_{i} \rangle$ in the vicinity of the $T_{c}$ follows an asymptotic form~\cite{fridkin2014,Fei2016} of $\langle P_{i} \rangle=C (T_{c}-T)^{\delta}$ with $T<T_{c}$. Here, $C$ is a constant and $\delta$ is the critical exponent. For monolayer GeTe, the asymptotic form fits well with the MC simulations, as shown in Fig.~\ref{wh3}(a). $P_{s}$ decreases continuously to zero at $T_{c}$. The $\delta$ is 0.195, deviating from $\delta$=0.5 in the second-order ferroelectric phase transition.~\cite{fridkin2014} A similar behavior has also been reported in other IV-VI compounds such as SnSe.~\cite{Fei2016}

Based on the Landau-Ginzburg phase transition theory,~\cite{Cowley1980} the electric field $E$ can be calculated from free energy, i.e. $E=\frac{\partial F(P)}{\partial P}$. The coercive field $E_{c}$ is at the turning points of the hysteresis loop $P(E)$, satisfying the condition of $(\frac{\partial P}{\partial E})^{-1}|_{E=E_{c}}=0$. Therefore, the ideal $E_{c}$ can be estimated from the maximum slope of the $F(P)$ curve between the NP and FE states.~\cite{Wang2016} A lateral size of $l=30$ nm is adopted to estimate the effective coercive voltage $V_{c}=lE_{c}$. This lateral size is about the one of latest ferroelectric field effect transistor memory.~\cite{Mueller2016} Through the $F(P)$ curve of monolayer GeTe (Fig.~\ref{wh2}(b)), the estimated $E_{c}$ is 0.206 V/nm and the effective $V_{c}$ is 6.18 V.
It is noted that the ideal $E_{c}$ of the bulk ferroelectric material is always much higher than the experimentally measured $E_{c}$, due to the growth and propagation of the ferroelectric domains.~\cite{Kim2002} However, the distinction between them at the nanoscale become small as thin ferroelectric films turn out to be more homogeneous than bulk and the formation of ferroelectric domains is suppressed.~\cite{Highland2010,fridkin2014}
\begin{figure}[tbp!]
\centerline{\includegraphics[width=0.45\textwidth]{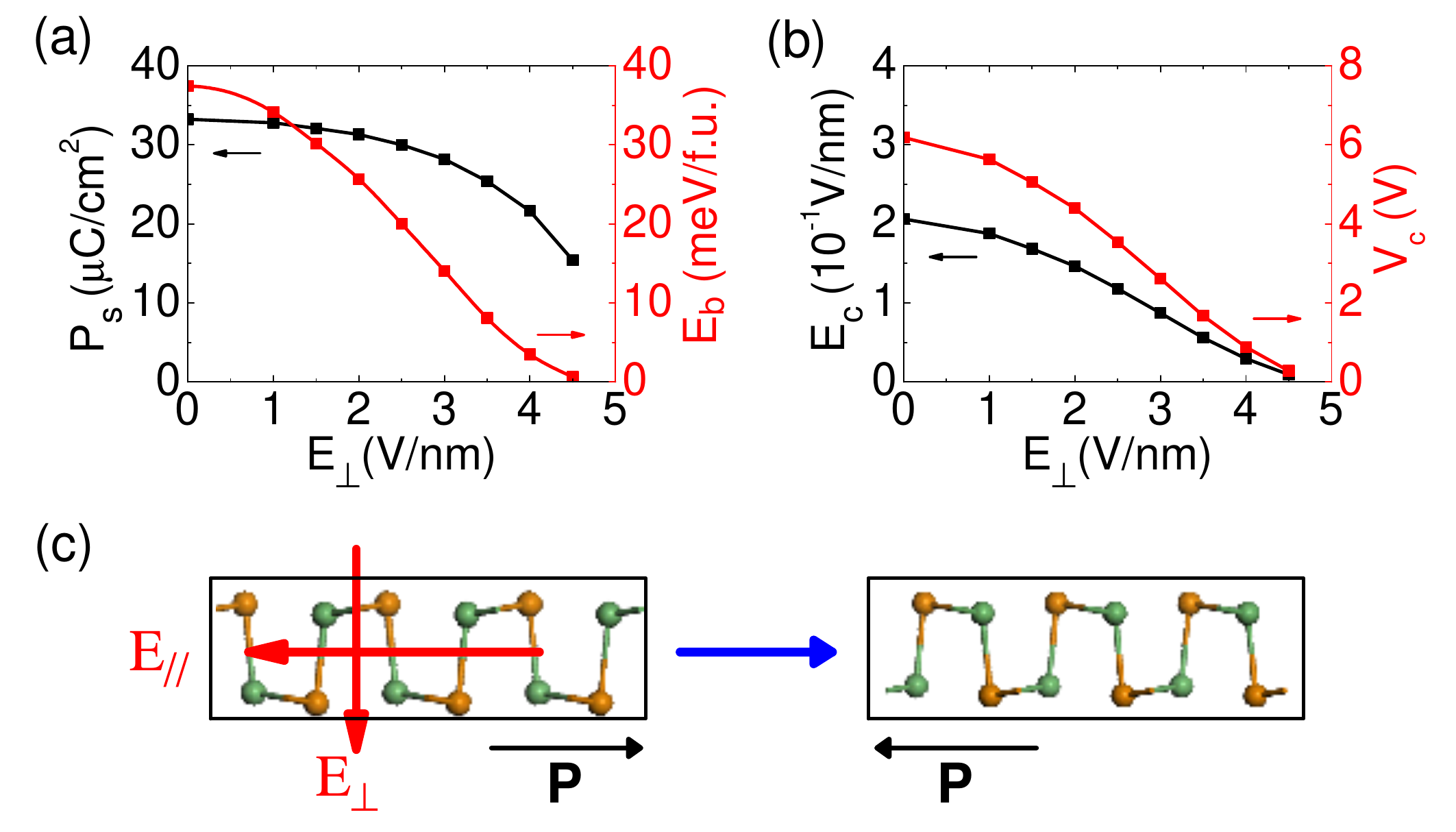}}
\caption{(a) Vertical electric field $E_{\perp}$ dependence of polarization $P_{s}$ and transition barrier $E_{b}$ of monolayer GeTe. (b) Vertical electric field $E_{\perp}$ dependence of in-plane coercive field $E_{c}$ and coercive voltage $V_{c}$ of monolayer GeTe. (c) A schematic representation of the switching of polarization of monolayer GeTe by a combination of an in-plane electric field $E_{//}$ and an out-of-plane electric field $E_{\perp}$.}
\label{wh4}
\end{figure}

It is found that if a vertical electric field $E_{\perp}$ was applied, the in-plane displacements of the Ge atoms with respect to the Te atoms will decrease, due to the field-induced coulomb forces. This leads to a reduction of the $P_{s}$ and $E_{b}$ (see Fig.~\ref{wh4}(a)). As a result, the in-plane coercive field $E_{c}$ in monolayer GeTe can be effectively decreased by $E_{\perp}$, as displayed in Fig.~\ref{wh4}(b). The maximum $E_{\perp}$ required to tune the $E_{c}$ is about 4.5 V/nm (see Fig.~\ref{wh4}(b). The equivalent $V_{\perp}$ is 4.5 V if an effective thickness of 1 nm for monolayer GeTe is adopted.
Fig.~\ref{wh4}(c) depicts a feasible way for fast switching the polarization by a combination of two mutually perpendicular electric fields. In this switching process, the required operating voltages are less than 5V, which is desirable for the integration into Si-based semiconducting devices.~\cite{Scott2007,Rabe2005}

\begin{figure}[tbp!]
\centerline{\includegraphics[width=0.45\textwidth]{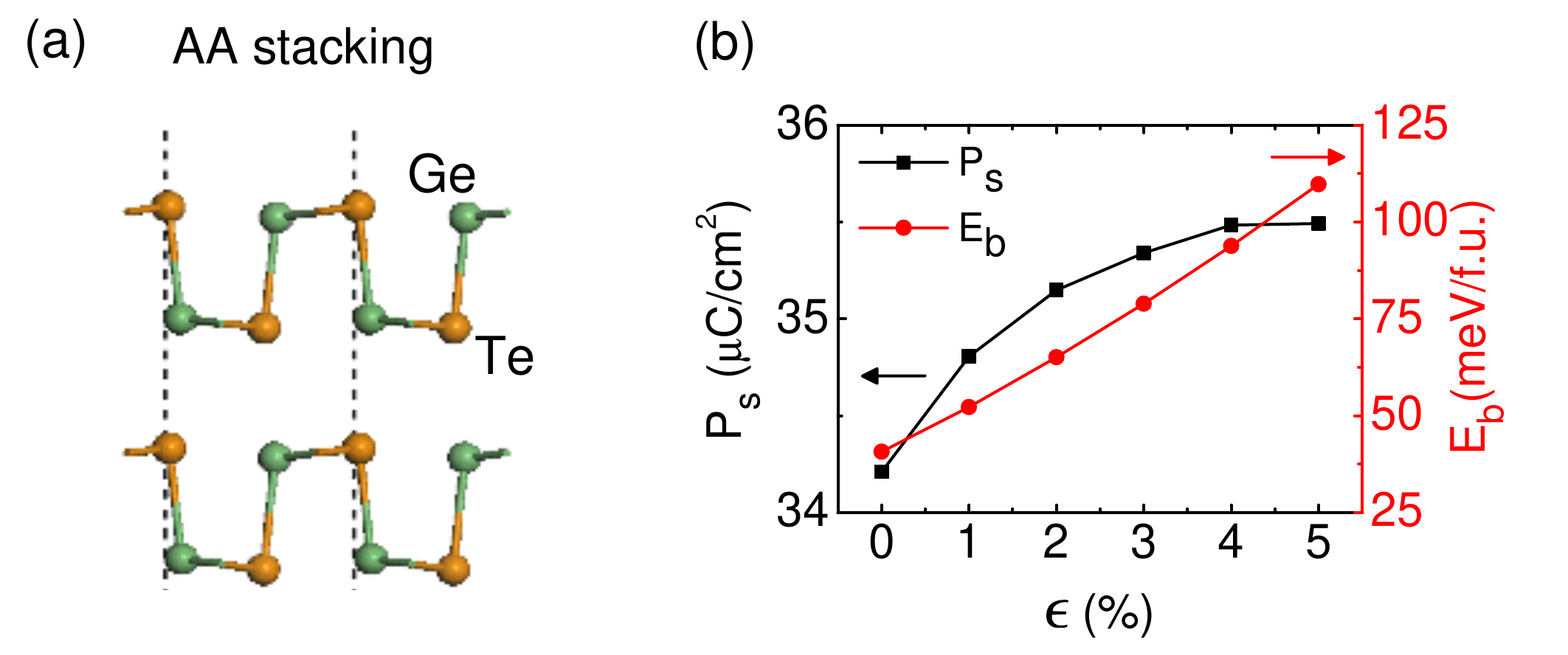}}
\caption{The (a) AA stacking order for bilayer GeTe. (b) Biaxial strain dependence of the $P_{s}$ and $E_{b}$ of bilayer GeTe with the AA stacking.}
\label{wh5}
\end{figure}

The stacking order is crucial for the ferroelectricity of multilayers of Group-IV tellurides. We first defined two kinds of stacking order for bilayer GeTe, namely, AA and anti-AA stacking. As shown in Fig.~\ref{wh5}(a), for AA-stacking, the top layer is directly stacked on top of the bottom layer, so that polarization of each layer is aligned parallel. By further shifting the top layer of AA-stacking with $\bm{a}$/2, $\bm{b}$/2 and $(\bm{a+b})$/2, we can get other three stacking orders labeled as AB, AC and AD stacking. The anti-AA stacking can be gotten if the top layer of the AA-stacked bilayer is rotated around the $z$ axis by $180^{\circ}$. Other stacking orders, e.g. anti-AB, anti-AC and anti-AD can be obtained with a similar process.
The energies of bilayer GeTe with different stacking order are shown in Table S4.~\cite{support} The bilayer GeTe with the AA stacking has the lowest energy.
The effective thickness of bilayer GeTe is taken as the twice of that of its monolayer. The corresponding effective bulk $P_{s}$ and $E_{b}$ is 34.2 ${\mu}$C/cm$^{2}$ and 40.1 meV/f.u., respectively, exhibiting a increase compared with that of monolayer GeTe (see Fig.~\ref{wh2}(c) and Fig.~\ref{wh5}(d)). Tensile strain can further enhance the $P_{s}$ and $E_{b}$ of bilayer GeTe (see Fig.~\ref{wh5}(d)). Thus, the promising ferroelectricity also exists in bilayer GeTe.

For bilayer SnTe, the AA-stacking is also the ground state but exhibits a weak ferroelectricity (see Table S4~\cite{support}).
Therefore, 2D GeTe and SnTe take the advantage over aforementioned Group-IV monochalcogenides such as SnSe in ferroelectric application, as the ferroelectricity in 2D SnSe only exists in odd-numbered layers.~\cite{Fei2016}

In summary, we show by the first-principles calculations that 2D GeTe with a hinge-like crystal structure is ferroelectric with an in-plane spontaneous polarization.
When examining the atomic structure and ferroelectricity of 2D GeTe, it is necessary to include of the van der Waals interactions in order to well describe the interatomic interactions.
The Curie temperatures $T_{c}$ of monolayer GeTe is as high as 570 K. Tensile strain can effectively enhance the $T_{c}$ and serves as a powerful tool to improve the ferroelectricity of 2D GeTe. The in-plane coercive field for reversing the polarization can be widely tuned by a vertical electric field, facilitating the fast switching of polarization. Furthermore, for bilayer GeTe the ferroelectric phase is still the ground state. With these advantages, 2D GeTe may be the long-sought candidate for realizing the integrated ferroelectric applications.

We acknowledge professor Wei Kang (Peking university) and Dr. Ruixiang Fei (Washington University) for their fruitful discussions. The work is supported by the National Natural Science Foundation of China (Grant No. 11574029) and the MOST Project of China (Nos. 2014CB920903, 2016YFA0300600).

\nocite{*}
\bibliography{GroupIVTellurides}
\end{document}